\documentclass{elsart5p}

\usepackage{graphics}

\journal{Physics Letters B}

\begin{document}

\renewcommand{\textfraction}{0.00000000001}
\renewcommand{\floatpagefraction}{1.0}
\begin{frontmatter}
\title{Measurement of the beam-helicity asymmetry $I^{\odot}$ in the  
photoproduction of $\pi^0$-pairs off the proton and off the neutron}

\author[Basel]{M.~Oberle},
\author[Basel]{B.~Krusche},\ead{Bernd.Krusche@unibas.ch}   
\author[Mainz]{J.~Ahrens},             
\author[Glasgow]{J.R.M.~Annand},          
\author[Mainz]{H.J.~Arends},
\author[Kent]{K.~Bantawa},
\author[Mainz]{P.A.~Bartolome},            
\author[Bonn]{R.~Beck},         
\author[Petersburg]{V.~Bekrenev},
\author[Giessen]{H.~Bergh\"auser},            
\author[Pavia]{A.~Braghieri},           
\author[Edinburgh]{D.~Branford},            
\author[Washington]{W.J.~Briscoe},           
\author[UCLA]{J.~Brudvik},             
\author[Lebedev]{S.~Cherepnya},           
\author[Washington]{B.~Demissie},
\author[Basel]{M.~Dieterle},             
\author[Mainz,Glasgow,Washington]{E.J.~Downie},            
\author[Giessen]{P.~Drexler}, 
\author[Lebedev]{L.V. Fil'kov}, 
\author[Tomsk]{A. Fix},             
\author[Edinburgh]{D.I.~Glazier},           
\author[Mainz]{E.~Heid}, 
\author[Sackville]{D.~Hornidge}, 
\author[Glasgow]{D.~Howdle}, 
\author[Regina]{G.M.~Huber},           
\author[Mainz]{O.~Jahn},
\author[Basel]{I.~Jaegle},
\author[Edinburgh]{T.C.~Jude},
\author[Basel]{A. K{\"a}ser},                
\author[Lebedev,Mainz]{V.L.~Kashevarov},
\author[Basel]{I. Keshelashvili},        
\author[Moskau]{R.~Kondratiev},          
\author[Zagreb]{M.~Korolija},  
\author[Petersburg]{S.P. Kruglov}, 
\author[Petersburg]{A.~Kulbardis},
\author[Moskau]{V.~Lisin},               
\author[Glasgow]{K.~Livingston},          
\author[Glasgow]{I.J.D.~MacGregor},       
\author[Basel]{Y.~Maghrbi},
\author[Glasgow]{J.~Mancell},  
\author[Kent]{D.M.~Manley}, 
\author[Washington]{Z.~Marinides},           
\author[Mainz]{M.~Martinez},
\author[Glasgow]{J.C.~McGeorge}, 
\author[Glasgow]{E.~McNicoll},          
\author[Zagreb]{D.~Mekterovic},          
\author[Giessen]{V.~Metag},
\author[Zagreb]{S.~Micanovic},
\author[Sackville]{D.G.~Middleton},
\author[Pavia]{A.~Mushkarenkov},               
\author[UCLA]{B.M.K.~Nefkens},         
\author[Bonn]{A.~Nikolaev},   
\author[Giessen]{R.~Novotny},
\author[Mainz]{M.~Ostrick},
\author[Mainz,Washington]{B.~Oussena},
\author[Pavia]{P.~Pedroni}, 
\author[Basel]{F.~Pheron},            
\author[Moskau]{A.~Polonski},            
\author[UCLA]{S.N.~Prakhov}, 
\author[Glasgow]{J.~Robinson},                    
\author[Glasgow]{G.~Rosner},              
\author[Basel,Pavia]{T.~Rostomyan},           
\author[Mainz]{S.~Schumann},
\author[Edinburgh]{M.H.~Sikora},            
\author[Catholic]{D.I.~Sober},               
\author[UCLA]{A.~Starostin},           
\author[Zagreb]{I.~Supek},               
\author[Giessen]{M.~Thiel},        
\author[Mainz]{A.~Thomas},              
\author[Mainz,Bonn]{M.~Unverzagt},                       
\author[Edinburgh]{D.P.~Watts},
\author[Basel]{D.~Werthm\"uller},
\author[Basel]{L.~Witthauer}, 
\author[Basel]{F.~Zehr}

\address[Basel] {Department of Physics, University of Basel, Ch-4056 Basel, Switzerland}
\address[Mainz] {Institut f\"ur Kernphysik, University of Mainz, D-55099 Mainz, Germany}
\address[Glasgow] {School of Physics and Astronomy, University of Glasgow, G12 8QQ, United Kingdom}
\address[Kent] {Kent State University, Kent, Ohio 44242, USA}
\address[Bonn] {Helmholtz-Institut f\"ur Strahlen- und Kernphysik, University Bonn, D-53115 Bonn, Germany}  
\address[Petersburg] {Petersburg Nuclear Physics Institute, RU-188300 Gatchina, Russia}
\address[Giessen] {II. Physikalisches Institut, University of Giessen, D-35392 Giessen, Germany}
\address[Pavia] {INFN Sezione di Pavia, I-27100 Pavia, Pavia, Italy}
\address[Edinburgh] {School of Physics, University of Edinburgh, Edinburgh EH9 3JZ, United Kingdom}
\address[Washington] {Center for Nuclear Studies, The George Washington University, Washington, DC 20052, USA}
\address[UCLA] {University of California Los Angeles, Los Angeles, California 90095-1547, USA}
\address[Lebedev] {Lebedev Physical Institute, RU-119991 Moscow, Russia}
\address[Tomsk] {Laboratory of Mathematical Physics, Tomsk Polytechnic University, Tomsk, Russia}
\address[Sackville] {Mount Allison University, Sackville, New Brunswick E4L1E6, Canada}
\address[Regina] {University of Regina, Regina, SK S4S-0A2 Canada}
\address[Moskau] {Institute for Nuclear Research, RU-125047 Moscow, Russia}
\address[Zagreb] {Rudjer Boskovic Institute, HR-10000 Zagreb, Croatia}
\address[Catholic] {The Catholic University of America, Washington, DC 20064, USA}

\begin{abstract}
Beam-helicity asymmetries have been measured at the MAMI accelerator in Mainz for
the photoproduction of neutral pion pairs in the reactions 
$\vec{\gamma}p\rightarrow p\pi^0\pi^0$ and $\vec{\gamma}d\rightarrow (n)p\pi^0\pi^0$,
$\vec{\gamma}d\rightarrow (p)n\pi^0\pi^0$ off free protons and off quasi-free 
nucleons bound in the deuteron for incident photon energies up to 1.4 GeV. 
Circularly polarized photons were produced from bremsstrahlung of longitudinally 
polarized electrons and tagged with the Glasgow magnetic spectrometer. Decay photons 
from the $\pi^0$ mesons, recoil protons, and recoil neutrons were detected in the 
4$\pi$ covering electromagnetic calorimeter composed of the Crystal Ball and TAPS 
detectors. After kinematic reconstruction of the final state, excellent agreement was 
found between the results for free and quasi-free protons. This demonstrates that the 
free-nucleon behavior of such observables can be extracted from measurements with 
quasi-free nucleons, which is the only possibility for the neutron. 
Contrary to expectations, the measured asymmetries are very similar for reactions 
off protons and neutrons. The results are compared to the predictions from the 
Two-Pion-MAID reaction model and (for the proton) also to the Bonn-Gatchina coupled 
channel analysis.   
  \end{abstract}
\end{frontmatter}

\section{Introduction}
Photoproduction of meson pairs has attracted much interest 
as a tool for the study of the excitation spectrum of the nucleon. 
In the absence of analytic or perturbative solutions of Quantum Chromodynamics
on the energy scale typical for excited states of the nucleon, up till now 
the experimental results have been mainly interpreted with phenomenological 
quark models. Although it is evident, that such models can only serve as an 
approximation of the complicated structure of the nucleon, a comparison to 
experiment should give some guidance concerning the relevant properties of 
the interaction. However, so far this approach has had only 
limited success. Apart from problems with the ordering of some low lying states,
already at moderate excitation energies the match in simple number counting 
between experiment and models is poor; many more states are predicted than have 
been observed  \cite{PDG,Krusche_03}. With a few exceptions, for most combinations 
of quantum numbers only the lowest lying state is known experimentally \cite{PDG}, 
while models predict a plethora of higher lying states. 

Theoretical developments are progressing mainly through advances in lattice 
gauge calculations and their combination with the methods of chiral perturbation 
theory for the extrapolation to physical quark masses. Interestingly, first 
unquenched lattice results for excited nucleon states \cite{Edwards_11}, basically 
`re-discovered' the well-known SU(6)$\otimes$O(3) excitation structure of the nucleon 
with a level counting consistent with the standard non-relativistic quark model. 
However, one should keep in mind that these calculations are still at a very early 
stage and are far from the quality already reached for ground-state properties 
\cite{Duerr_08}.  

On the experimental side the data base may be biased because it was dominated 
by pion-scattering so that partial wave analyses like \cite{Arndt_06} may be 
insensitive to resonances weakly coupled to the $N\pi$ channel. This has prompted 
world-wide efforts to study excited nucleon states with photon-induced
meson-production reactions. Photoproduction of meson pairs adds an important component 
to these efforts. It allows the study of resonances that have no significant direct 
decay mode to the nucleon ground state via single meson production. Such states decay
via intermediate excited states in sequential reaction chains such as 
$R\rightarrow R'\pi\rightarrow N\pi\pi$ ($R$, $R'$ nucleon resonances, $N=n,p$). 
It is obvious that for higher lying resonances such processes will be important. 
As an analogy, in nuclear physics a restriction to ground-state decays of excited states 
would result in a very rudimentary picture of nuclear structure.

For the photoproduction of pion pairs data covering the second resonance region,
which includes the $P_{11}$(1440), $S_{11}$(1535), and $D_{13}$(1520) states,
has been collected with photons of energies up to $\approx$0.9 GeV.
A smaller number of measurements have been made at higher incident photon energies up to 
1.5~GeV. Total cross sections, invariant-mass distributions of the $\pi\pi$- and the 
$\pi N$-pairs, and angular distributions have been measured with the DAPHNE, TAPS, and 
Crystal Ball detectors at the MAMI accelerator in Mainz 
\cite{Braghieri_95,Haerter_97,Zabrodin_97,Zabrodin_99,Wolf_00,Kleber_00,Langgaertner_01,Kotulla_04,Sarantsev_08,Zehr_12,Kashevarov_12},
at GRAAL in Grenoble \cite{Assafiri_03,Ajaka_07},
at ELSA in Bonn \cite{Sarantsev_08,Thoma_08}, and with the CLAS detector at JLab 
(with electron beams) \cite{Ripani_03}. Recently, first polarization observables were also 
measured at the MAMI accelerator \cite{Ahrens_03,Ahrens_05,Ahrens_07,Ahrens_11}.
A discussion of the data for the $p\pi^0\pi^0$ final state in terms of nucleon 
resonances is given in \cite{Thoma_08} in the framework of the Bonn-Gatchina (BnGa) 
coupled channel analysis. Double $\pi^0$ production is particularly interesting 
since it has much smaller contributions from non-resonant background processes like
pion-pole diagrams, or terms of the $\Delta$-Kroll Rudermann type, than the channels
with charged pions.

However, the photoproduction of pseudo-scalar meson pairs off nucleons \cite{Roberts_05} is described 
by eight complex amplitudes as function of five kinematic observables (for example two 
Lorentz invariants and three angles). Measurements of eight independent observables are required (the helicity asymmetry $I^{\odot}$ discussed below can be chosen as one 
of them) to extract the magnitudes of those amplitudes. Fixing in addition 
all relative phases requires in total the measurement of at least 15 
observables (single, double, and triple polarization observables among them), all 
as function of the five kinematic variables. Due to finite statistical precision
of the results, in practice even more observables would have to be measured, which
is not realistic.

Therefore, reaction models are used to extract the most relevant information about 
nucleon resonances from the measurements. They use parametrizations 
for the resonance and background contributions and sometimes coupled channel techniques 
to combine the results from many meson production reactions, so that the fits 
become better constraint. However, even in such a framework results 
for at least some polarization observables are indispensable.

The results from such model analyses for double $\pi^0$ production are still
controversial even at low excitation energies, where only a few resonances 
contribute. Different reaction models propose, for example, dominant contributions of the 
$D_{13}(1520)\rightarrow \Delta\pi^0\rightarrow p\pi^0\pi^0$ reaction chain
\cite{Gomez_96,Nacher_01,Nacher_02,Fix_05}, or of the  
$P_{11}(1440)\rightarrow N\sigma$ decay \cite{Assafiri_03}.
The BnGa analysis \cite{Sarantsev_08,Thoma_08} finds a large contribution from the 
$D_{33}(1700)$ state, which is almost negligible in the other models.
The helicity dependence of the cross sections, measured in context with the 
Gerasimov-Drell-Hearn sum rule, \cite{Ahrens_03,Ahrens_05,Ahrens_07} shows a 
dominance of the $\sigma_{3/2}$ component. This is in line with the excitation of 
the $D_{13}$ or $D_{33}$ states and limits possible $P_{11}$ contributions. 
Very recently, Kashevarov et al. \cite{Kashevarov_12} have presented a detailed analysis 
of the angular distributions. They argue that in the $D_{13}$(1520) range, and even below, 
a large contribution of $J=3/2$ waves is necessary to reproduce the data. 
However, they cannot identify the nature of this contribution. Possible candidates are a 
strong D$_{33}$(1700) excitation (which, however, is probably in conflict with other 
isospin channels) or $\pi^+\pi^-\rightarrow \pi^0\pi^0$ rescattering effects, which are 
neglected in most models but may play an important role.

\section{Beam-helicity asymmetries}
\begin{figure}[htb]
\resizebox{0.49\textwidth}{!}{%
  \includegraphics{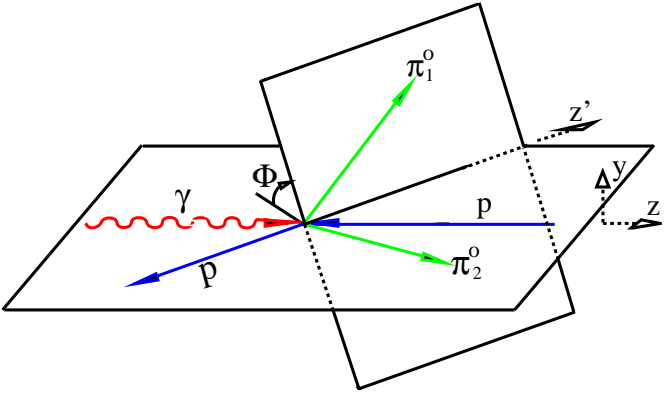}
}
\caption{Vector and angle definitions in the cm system. $\Phi$ is the angle 
between the reaction plane (defined by $\vec{k}$ and $\vec{p}_{p}$) and the 
production plane of the two pions (defined by
$\vec{p}_{\pi^0_1}$ and $\vec{p}_{\pi^0_2}$).
}
\label{fig:def}       
\end{figure}

An observable that recently attracted much interest is the beam-helicity asymmetry
$I^{\odot}$ \cite{Nacher_02,Fix_05,Roca_05}. It can be measured for three-body final 
states such as $N\pi\pi$ with circularly polarized photons and unpolarized targets. 
Models  \cite{Roca_05} predicted a large sensitivity to different reaction mechanisms. 
This observable is defined by:
\begin{equation}
I^{\odot}(\Phi)=\frac{1}{P_{\gamma}}
	        \frac{d\sigma^{+}-d\sigma^{-}}{d\sigma^{+}+d\sigma^{-}}
	       =\frac{1}{P_{\gamma}}
                \frac{N^{+}-N^{-}}{N^{+}+N^{-}}\; ,
\label{eq:circ}		
\end{equation}
where $d\sigma^{\pm}$ are the differential cross sections for each of the two
photon helicity states, and $P_{\gamma}$ is the degree of circular polarization 
of the photons. The definition of the angle $\Phi$ \cite{Roca_05} in the 
center-of-momentum (cm) system between the reaction plane, spanned by the incident 
photon and the outgoing recoil nucleon, and the production plane, spanned by the 
two pions, is illustrated in Fig. \ref{fig:def}. Note that $\Phi$ depends on
the numbering of the two pions.
Parity conservation imposes on the asymmetry the condition 
\begin{equation}
I^{\odot}(\Phi)=-I^{\odot}(2\pi-\Phi)
\label{eq:sym1}
\end{equation}
and for identical mesons, like $\pi^0\pi^0$ pairs, when the numbering of the 
two pions is randomized, it must obey in addition 
\begin{equation}
I^{\odot}(\Phi)=I^{\odot}(\Phi +\pi).
\label{eq:sym2}
\end{equation}
This relation does not apply when the two pions are ordered by some
kinematic condition, for example by the invariant masses $m(\pi^0,N)$
of the pion-nucleon pairs
\begin{equation}
m(\pi^0_1,N)\ge m(\pi^0_2,N)\;,
\label{eq:mmcon}
\end{equation}
selecting the pion with the larger invariant mass as number one.
For the extraction of $I^{\odot}(E_{\gamma},\Phi,\Theta_{\pi_1},\Theta_{\pi_2},...)$ 
in a limited region of kinematics, the differential cross sections $d\sigma^{\pm}$ 
can be replaced by the respective count rates $N^{\pm}$, since all normalization factors 
cancel in the ratio. For integrated asymmetries, efficiency weighted count rates 
$N/\epsilon$ must be used.  

Previously, this asymmetry has been measured for double pion production for the
$\vec{\gamma} p\rightarrow p\pi^+\pi^-$ reaction with CLAS at JLab up to incident photon
energies of 2.3~GeV \cite{Strauch_05} and for all three isospin states 
($\pi^0\pi^0$, $\pi^+\pi^-$, $\pi^0\pi^+$) for the $\vec{\gamma} p$ initial state at MAMI 
up to photon energies of 0.8~GeV \cite{Krambrich_09}. These studies found that
states are more likelynone of the available reaction models agreed with the asymmetries
measured for the reactions involving charged meson production. 
Only for the double $\pi^0$ channel, which has only small non-resonant contributions, 
reasonable agreement was found \cite{Krambrich_09} for the models of Fix and Arenh\"ovel 
\cite{Fix_05} and the BnGa analysis \cite{Sarantsev_08}; however, the Valencia model 
\cite{Roca_05} failed although it agreed very well with the total cross section and 
invariant mass distributions. This demonstrates the impact of such data.

The present experiment concentrates on the neutral channel, extending the 
measured energy range for $I^{\odot}$ for $\vec{\gamma} p\rightarrow p\pi^0\pi^0$ to 1.4 GeV, 
throughout the third resonance region. At the same time it provides the first data 
for $I^{\odot}$ for the $\vec{\gamma} n\rightarrow n\pi^0\pi^0$ reaction. The study of
photon-induced reactions off the neutron has recently moved more into focus. It can yield 
valuable complementary information since the photoexcitation of nucleon resonances is 
isospin dependent and some states are more likely excited for the neutron than for the 
proton (see \cite{Krusche_11} for an overview over this program).  
 
\section{Experiment and analysis}
The measurements were performed at the tagged photon beam of the Mainz MAMI accelerator 
\cite{Herminghaus_83,Kaiser_08}. Longitudinally polarized electron beams with energies of 
$\approx$1.5 GeV (see Table. \ref{tab:beam} for details) were used to produce bremsstrahlung
photons in a copper radiator of 10 $\mu$m thickness. The photons were tagged with the upgraded 
Glasgow magnetic spectrometer \cite{Anthony_91,Hall_96,McGeorge_08}. The typical bin width 
for the photon beam energy (4 MeV) was defined by the geometrical size of the plastic 
scintillators in the focal plane detector of the tagger. 
The polarization state of the beam was switched in a randomized way with a frequency of 1 Hz.
Possible differences in the number of incident photons for the two helicity states have 
been determined to be at the 5$\times$10$^{-4}$ level, i.e. they are negligible.
The polarization degree of the electron beams was measured by Mott and M$\o$ller scattering. 
Their longitudinal polarization is transferred in the bremsstrahlung process to circular 
polarization of the photons. The polarization degree of the photon beams follows from the 
polarization degree of the electrons and the energy-dependent polarization transfer factors 
given by Olsen and Maximon \cite{Olsen_59}. 

The size of the tagged photon beam spot on the targets ($\approx$ 1.3 cm diameter) was 
restricted by a collimator (4~mm diameter) placed downstream from the radiator foil. 
The targets were Kapton cylinders of $\approx$ 4 cm diameter and different lengths filled 
with liquid deuterium or liquid hydrogen. Contributions from the target windows 
(2$\times$120~$\mu$m Kapton) were determined with empty target measurements, but are negligible 
for the results discussed in this paper. Data were taken during four different beam times. 
Their main parameters are summarized in Table~\ref{tab:beam}.

\begin{table}[hhh]
\begin{center}
  \caption[Summary of data sets]{
    \label{tab:beam}
     Summary of data samples. 
     Target type ($LD_2$: liquid deuterium, $\rho_d$ = 0.169 g/cm$^3$; 
     $LH_2$: liquid hydrogen, $\rho_H$ = 0.071 g/cm$^3$), target length 
     $d$ [cm], target surface density $\rho_s$ [nuclei/barn], 
     electron beam energy $E_{e^-}$ [MeV],
     degree of longitudinal polarization of electron beam $P_{e^-}$ [\%].
}
\vspace*{0.3cm}
\begin{tabular}{|c|c|c|c|c|c|}
\hline
~Data Set~ & ~Target~ & ~d [cm]~ & ~$\rho_s$ [barn$^{-1}]~$ & 
~$E_{e^-}$ [MeV]~ & ~$P_{e^-}$ [\%]~\\
\hline\hline
 I & $LD_2$ & 4.72 & 0.231$\pm$0.005 & 1508 & 61$\pm$4\\
 II & $LD_2$ & 4.72 & 0.231$\pm$0.005 & 1508 & 84.5$\pm$6\\
 III & $LD_2$ & 3.00 & 0.147$\pm$0.003 & 1557 & 75.5$\pm$4\\
 IV & $LH_2$ & 10.0 & 0.422$\pm$0.008 & 1557 & 75.5$\pm$4\\  
\hline
\end{tabular}
\end{center}
\end{table}

Photons, charged pions, and recoil nucleons produced in the target were detected with an
almost $4\pi$ covering electromagnetic calorimeter. It combined the Crystal Ball detector 
(CB) \cite{Starostin_01} with the TAPS detector \cite{Novotny_91,Gabler_94}. The CB is made 
of 672 NaI crystals and covered the full azimuthal range for polar angles from 20$^{\circ}$ 
to 160$^{\circ}$, corresponding to 93\% of the full solid angle. The TAPS detector, consisting
of 384 BaF$_2$ crystals, was configured as a forward wall, placed 1.457 m downstream from the
targets, and covered polar angles from $\approx$5$^{\circ}$ to 21$^{\circ}$.
The Crystal Ball was equipped with an additional Particle Identification 
Detector (PID) \cite{Watts_04} for the identification of charged particles and 
all modules of the TAPS detector had individual plastic scintillators in front for the 
same purpose (TAPS `Veto-detector'). This setup is similar to the one described in detail in 
\cite{Zehr_12,Schumann_10}. 
The only difference is that, in the earlier setup, the TAPS forward wall consisted of 
510 modules and was placed 1.75~m from the target. The trigger conditions used for
the different beam times varied in the required multiplicity of hits in
the calorimeter (between 1 - 3). This is, however, irrelevant for the present analysis
since four-photon events from the $\pi^0\pi^0$ final state always fulfilled the
trigger conditions. It had only influence on the experimental dead time, limiting counting
statistics for the low-multiplicity runs. 
  
The analysis of double pion production reactions, measured with this setup, is discussed 
in detail in \cite{Zehr_12}. In the analysis, events were included with five neutral
hits or four neutral and one charged hit, where `neutral' and `charged' were defined by 
the response of the PID or the TAPS plastic scintillator system. They were taken as 
candidates for the $p\pi^0\pi^0$ and $n\pi^0\pi^0$ final states (the second event type
contributes only for the deuterium target). The analysis proceeded 
then with the identification of photons. For the TAPS detector, photons can be cleanly
separated from charged pions and recoil nucleons with the help of the charged particle
detectors, a pulse-shape analysis, and a time-of-flight-versus-energy analysis (see
\cite{Zehr_12} for details). For the CB, the separation was done with the PID.
Charged pions and protons were identified by a $\Delta E-E$ analysis of the signals from the
PID and the CB. Distinction between photons and neutrons is not possible for hits in the CB.
In the next step, identified photons (from TAPS) and photon candidates (from the CB) 
were combined into all possible disjunct pairs and their invariant masses were calculated. 
The `best' combination was chosen with a $\chi^2$ test minimizing the deviation of
the invariant masses from the nominal rest mass of the $\pi^0$. For events with
five photon candidates, the remaining bachelor was taken as a neutron candidate. 

\begin{figure}[htb]
\resizebox{0.48\textwidth}{!}{%
    \includegraphics{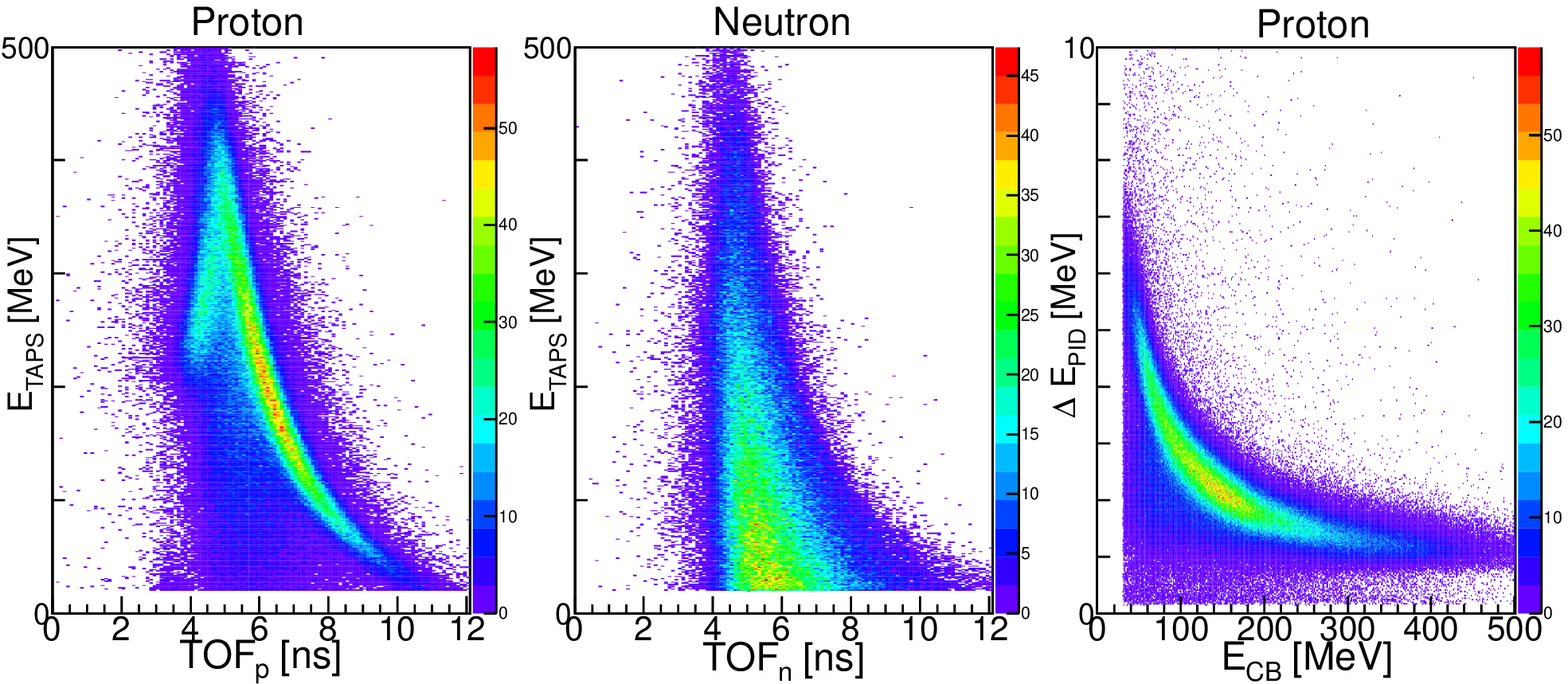}
}
\resizebox{0.30\textwidth}{!}{%
    \includegraphics{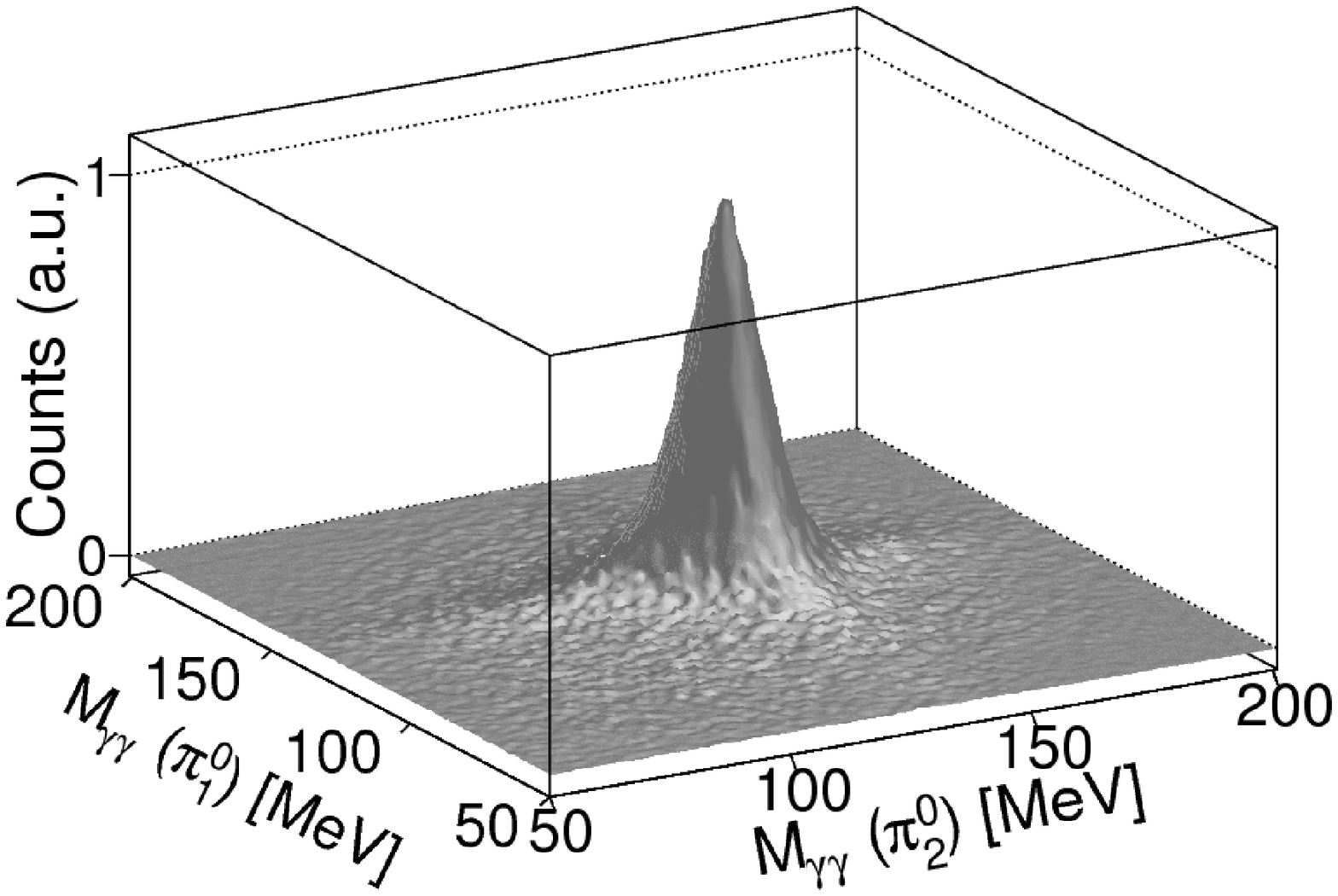}}
\hspace*{-0.4cm}\resizebox{0.20\textwidth}{!}{%
    \includegraphics{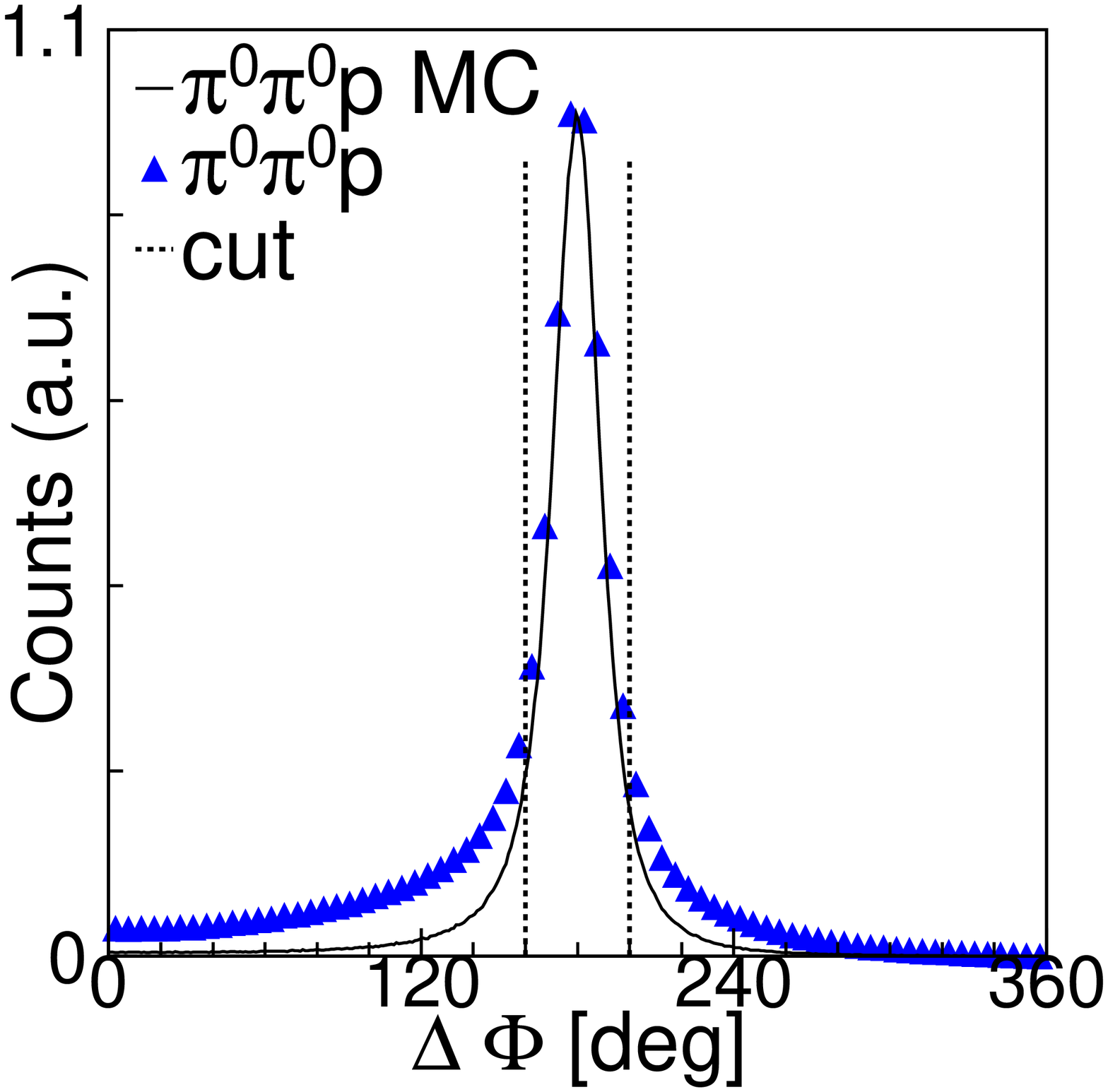}  
}  
\caption{Identification of particles. Top row, left hand side: ToF-versus-energy for 
proton candidates in TAPS,
center: same for neutrons, right hand side: $\Delta E-E$ spectrum for proton candidates 
in the CB. Bottom, left hand side: invariant masses of the `best' combination of four 
photons to two pairs for the $\vec{\gamma} N\rightarrow N\pi^0\pi^0$ reaction. 
Right hand side: co-planarity spectrum for events with recoil protons compared to MC 
simulation. All spectra for deuteron target.
}
\label{fig:invm}       
\end{figure}

The two-dimensional invariant-mass spectrum of the photon pairs from the `best'
solutions is shown in Fig.~\ref{fig:invm} (bottom left). The background level was 
very low already after these initial analysis steps. Only events with both pion invariant 
masses between 110 MeV and 160 MeV were accepted for further analysis and the nominal 
mass of the pion was used to improve the resolution. Since the angular resolution of 
the detector system is much better than the energy resolution, this was simply done 
by replacing the measured photon 
energies by
\begin{equation}
\label{eq:xform}
E'_{1,2}=E_{1,2}\frac{m_{\pi^0}}{m_{\gamma\gamma}}\;\;,
\end{equation}   
where $E_{1,2}$ are the  measured photon energies, $E'_{1,2}$ the re-calculated
energies, $m_{\pi^0}$ is the nominal $\pi^0$ mass, and $m_{\gamma\gamma}$ the
measured invariant mass. 

In the next step, the reaction kinematics in the cm-system was exploited by a selection 
on the co-planarity of the two pions with the recoil nucleon. The azimuthal angle between 
the cm three-vectors of the recoil nucleon and the combined two-pion system was required 
to be in the interval (180$\pm$20)$^{\circ}$ (see Fig.~\ref{fig:invm}, bottom right). 
Finally, the recoil nucleons (although detected) were treated as missing particles and 
their mass was computed from 
the reaction kinematic defining the missing mass $\Delta m$ by
\begin{equation}
\label{eq:2pimiss}
\Delta m(\pi\pi) = \left|P_{\gamma}+P_{N}- P_{\pi^0_1}
-P_{\pi^0_2}\right|
-m_N\ ,
\end{equation}
where $m_N$ is the nucleon mass, $P_{\gamma}$, $P_{N}$, $P_{\pi^0_{1,2}}$
are the four-momenta of the incident photon, the initial-state nucleon 
(which was assumed at rest, neglecting Fermi motion), and the produced $\pi^0$-mesons.
\begin{figure}[htb]
\resizebox{0.49\textwidth}{!}{%
  \includegraphics{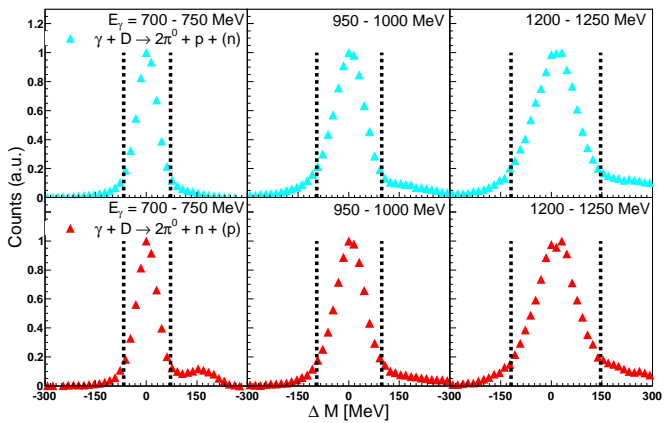}
}
\caption{Missing mass spectra for some typical incident photon energies for quasi-free
protons (upper part) and quasi-free neutrons (bottom part). Vertical, dashed lines:
applied cuts.
}
\label{fig:missm}       
\end{figure}
Typical results for double $\pi^0$ production off quasi-free nucleons 
from the deuterium target are shown in Fig.~\ref{fig:missm}. The spectra are rather clean,
the small background components for recoil neutrons are presumably due to the reaction 
$\gamma N\rightarrow N\eta$; $\eta\rightarrow 3\pi^0$, where one decay photon escaped 
detection and another one was misidentified as a neutron.

The correct identification of recoil nucleons is demonstrated in Fig.~\ref{fig:invm},
(top row), where the time-of-flight versus energy spectra of neutron and proton 
candidates in TAPS and the $\Delta E-E$ spectra for proton candidates in the CB are summarized.
Due to the careful selection of the different reaction types discussed above, these 
spectra are very clean. The spectra for charged recoil nucleons show no trace of events 
from charged pions. The ToF-versus-energy correlation for neutron candidates in TAPS 
does not show any residue from the proton band. Altogether, no indication of any significant
contamination of the proton sample with neutrons or vice versa has been observed.       

The asymmetries were constructed from these data samples using Eq.~\ref{eq:circ}.
Overall normalization factors (beam flux, target thickness, experimental deadtime) 
cancel in the ratio. However, in general, the experimental detection efficiency depends
on the reaction kinematics (polar angles and kinetic energies of pions and
recoil nucleons) so that it does not cancel in ratios integrated over such 
observables. This is not much of a problem for the detection of the $\pi^0$ mesons
because the detection efficiency of a $4\pi$ calorimeter is rather flat for them.
However, the detection probability for the recoil nucleons is sensitive to their kinetic 
laboratory energies and thus also to their cm polar angles (for fixed incident photon 
energy recoil energy drops with increasing polar angle). 
Therefore, the count rates $N^{\pm}$ were replaced by $N^{\pm}/\epsilon$, where 
$\epsilon$ is the experimental detection efficiency. The latter was constructed 
by Monte Carlo simulations with the Geant4 code \cite{Geant} and applied as function 
of $W$ ($\sqrt{s}$ of photon, participant-nucleon pair), $\Phi$ (cf. Fig.~\ref{fig:def} 
for definition of $\Phi$), and the recoil nucleon cm polar angle $\Theta_{N}^{\star}$. 
The other kinematical observables were integrated out, using a phase-space distribution 
for the $N\pi^0\pi^0$  final state in the event generator. This assumption is not 
critical since the detection efficiency is flat as function of these observables and 
does not much depend on the details of the event generator (see also discussion in 
\cite{Zehr_12}). Typical values for the detection efficiency for the reaction with 
recoil protons were around 20\%, but dropped to a few per cent for large polar angles 
($\Theta_{p}^{\star}>150^{\circ}$). Efficiencies around 10\% were reached with recoil
neutrons. The systematic uncertainty of the efficiency correction
is small, we estimate it below the 5\% range. The only further source of systematic 
uncertainty are the polarization degrees of the photon beam (see Table~\ref{tab:beam}),
which were measured with uncertainties between 5\% and 7\%.

For the quasi-free measurements, the effects of nuclear Fermi motion of nucleons bound 
in the deuteron must be considered. They will tend to smear out the 
asymmetries. However, as discussed in detail in \cite{Krusche_11,Jaegle_11}, the 
kinematics of the reaction is completely determined from the measurement of the 
incident photon energy, the four-vectors of the mesons, and the direction of 
the recoil nucleon. It is not necessary to measure directly the kinetic energy of the 
recoil nucleon. This is important because kinetic energies of recoil neutrons detected 
in the CB are not available (time-of-flight distance is too short for reasonable 
resolution). The four-vector of the participant (and also the spectator nucleon) 
can be reconstructed from this input so that the `true' $W$  and the correct cm-system 
of the photon - participant-nucleon pair are known. In all further analysis steps,
final state four-vectors reconstructed this way were used for recoil neutrons and
recoil protons.
The validity of this method can be tested by a comparison of the asymmetries extracted for
free protons from the hydrogen target to the results for quasi-free protons from the
deuterium target. 

\section{Results}
\begin{figure}[htb]
\resizebox{0.49\textwidth}{!}{%
  \includegraphics{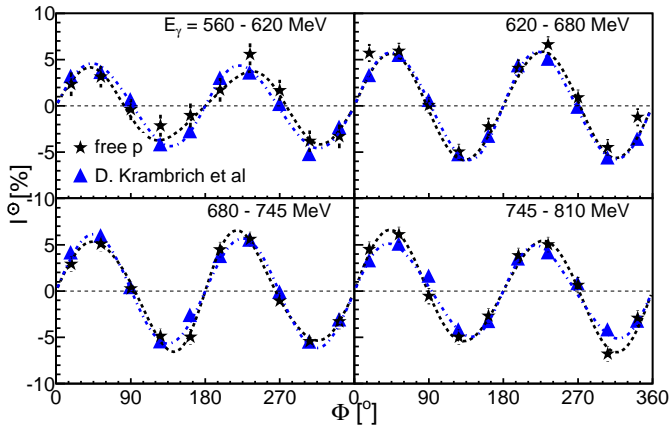}
}
\caption{Asymmetries for the free proton (black stars present experiment)
compared to previous results (magenta triangles) \cite{Krambrich_09}.
Dotted curves: fits to the data with Eq. \ref{eq:coeff}.  
}
\label{fig:zehr}       
\end{figure}

The asymmetries have been analyzed in two different ways, for a randomized
numbering of the two pions and for an ordering with the condition from Eq. \ref{eq:mmcon}.
The first analysis was only used for a comparison to the previous results from
\cite{Krambrich_09} for the free proton. This is shown in Fig.~\ref{fig:zehr}.
The agreement between the two measurements is excellent. 

\begin{figure*}[htb]
\centerline{\resizebox{1.0\textwidth}{!}{%
  \includegraphics{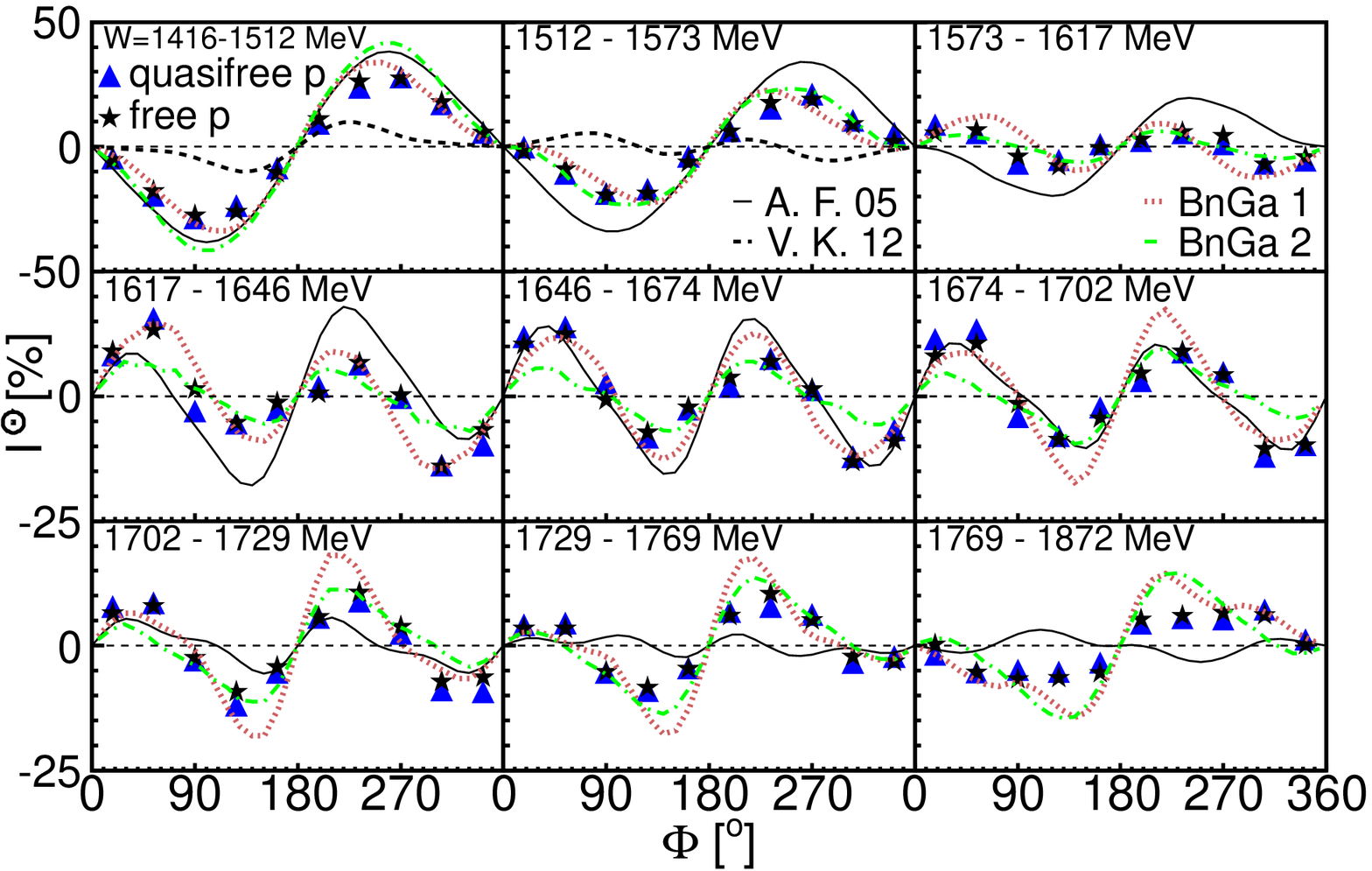}
  \includegraphics{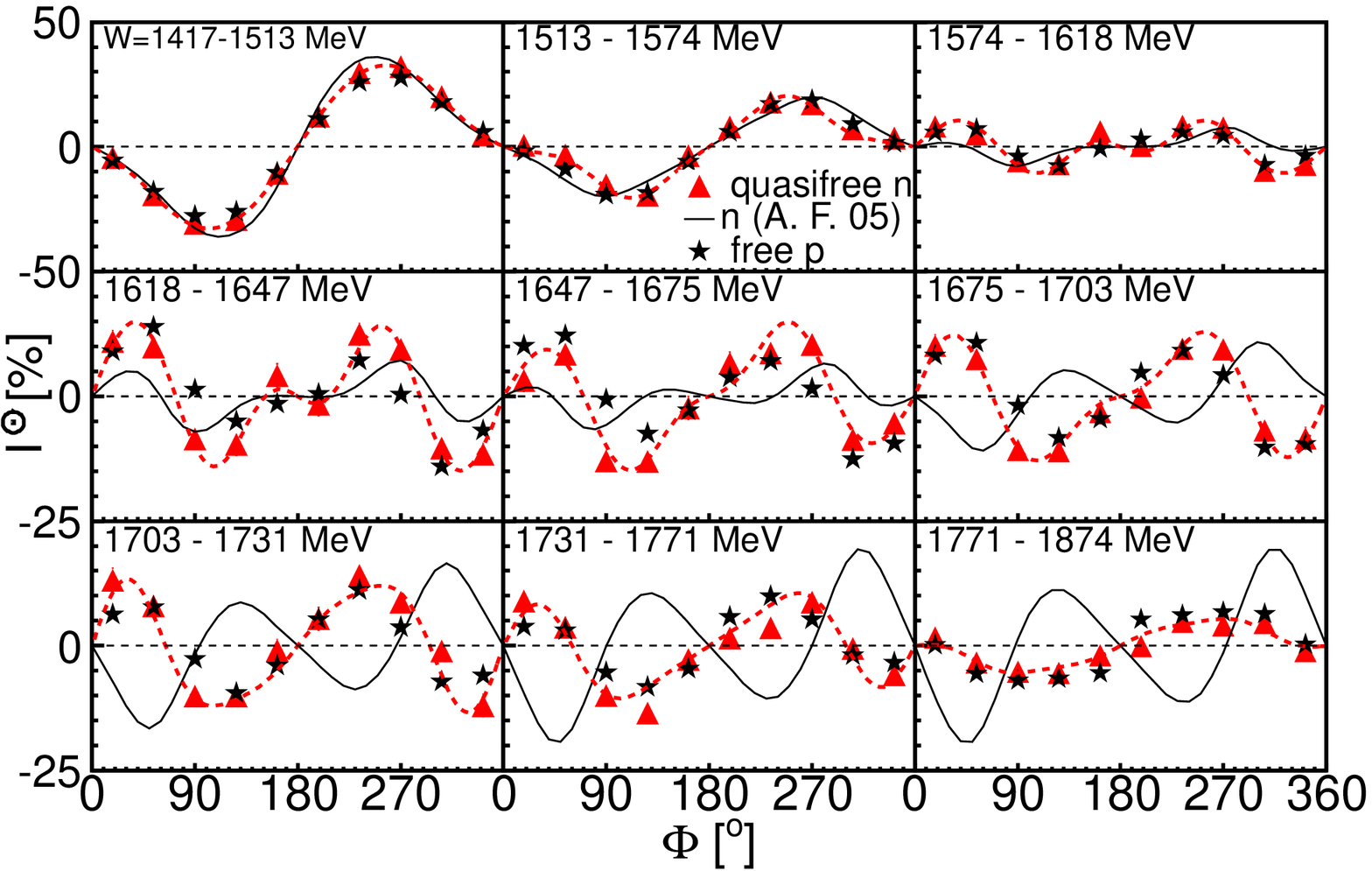}
}}
\caption{Results for $I^{\odot}(\Phi)$ for different ranges of $W=\sqrt{s}$. 
Left hand side: (black) stars: free proton, (blue) triangles: quasi-free proton.
(Black) solid curves (A.F. 05): model results from \cite{Fix_05}.
(Black) dotted curves (only for two lowest $W$ ranges): model results from 
\cite{Kashevarov_12} (V.K. 12).
(Magenta) dotted curves: solution BnGa2011-1 from \cite{Anisovich_12}, 
(green) dash-dotted curves: solution BnGa2011-2 from \cite{Anisovich_12}. 
Right hand side: data for quasi-free neutrons (red) triangles compared to free proton.
Dashed (red) curves: fits to neutron data. Solid (black) curves: model results
from \cite{Fix_05}. 
}
\label{fig:integ_asym}       
\end{figure*}

\begin{figure}[htb]
\resizebox{0.45\textwidth}{!}{%
  \includegraphics{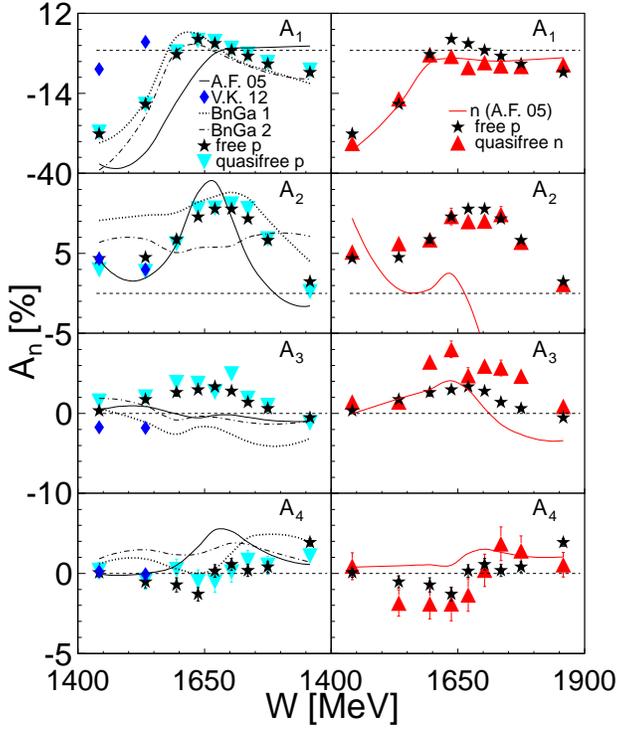}
}
\caption{Coefficients of the fits of the asymmetries from Fig.~\ref{fig:integ_asym}
with Eq.~\ref{eq:coeff} as a function of cm-energy $W$. Left-hand side: free and quasi-free 
proton data, right-hand side: comparison of proton and neutron data. 
Solid curves: model results from \cite{Fix_05}. Dotted curves: solution BnGa2011-1
from \cite{Anisovich_12}, dash-dotted curves: solution BnGa2011-2 from \cite{Anisovich_12}.
(Blue) diamonds (only for low energy region proton data): model results from 
\cite{Kashevarov_12}.
}
\label{fig:nucl_par}       
\end{figure}

The results from the second analysis carry more information. They have been 
fitted with a sine-series, reflecting the symmetry properties of this observable
as specified in Eq.~\ref{eq:sym1}: 
\begin{equation}
I^{\odot}(\Phi)=\sum_{n=1}^{\infty}A_{n}\mbox{sin}(n\Phi)\;.
\label{eq:coeff}
\end{equation}
For the ordered pion pairs, Eq.~\ref{eq:sym2} is not 
valid and consequently also the coefficients with odd numbers can contribute. The 
coefficients with even numbers are of course identical (within statistical 
fluctuations) to the corresponding coefficients of the analysis with randomized 
pions. The asymmetries are shown in Fig.~\ref{fig:integ_asym}, the coefficients are 
summarized in Fig.~\ref{fig:nucl_par}.

The asymmetries for the free and quasi-free proton are compared at the left hand side of
Fig.~\ref{fig:integ_asym} 
and their sine-coefficients are shown at the left hand side from Fig.~\ref{fig:nucl_par}. 
The comparison does not reveal any significant discrepancies, demonstrating
that the removal of effects from nuclear Fermi motion via the kinematic reconstruction 
works excellently. Thus, the neutron asymmetries measured this way can be taken as
a good approximation of free-neutron data and can be compared to corresponding model 
results. 

The comparison of the results for $\pi^0\pi^0$ production off protons and neutrons in
Figs.~\ref{fig:integ_asym} and \ref{fig:nucl_par} shows, over the entire 
investigated energy range, a surprising similarity between the asymmetries of the two 
isospin channels.
At low incident photon energies, in the so-called second resonance region, one might argue that 
the reaction mechanism for double $\pi^0$ production is similar for protons and neutrons, because 
in both cases the sequential $D_{13}(1525)\rightarrow \pi^0\Delta(1232)\rightarrow \pi^0\pi^0 N$
decay chain dominates. However, in the third resonance peak, around $W$=1.7 GeV, reaction models
\cite{Fix_05} predict dominant contributions from different resonances. Due to the electromagnetic 
$\gamma NN^{\star}$ couplings, the largest contribution to the proton cross section is supposed
to come from the $F_{15}(1680)\rightarrow \Delta\pi^0$ decay; while for the neutron the 
$D_{15}(1675)\rightarrow \Delta\pi^0$ decay is stronger \cite{Fix_05}. However, the observed
asymmetries are practically identical. Given the predicted large sensitivity of this observable 
to even small changes in the amplitudes (see for example \cite{Roca_05}), this stability is 
unexpected. 

Predictions for the asymmetries have been made by several reaction models. It has already been
shown in \cite{Krambrich_09} that the Valencia model \cite{Gomez_96,Nacher_01,Nacher_02,Roca_05}
does not reproduce the asymmetries in the second resonance region. Results at higher incident 
photon energies are not available from this model.

The Bonn-Gatchina coupled channel analysis has extracted in a recent update 
properties of baryon resonances from a simultaneous analysis of many different reaction
channels including double $\pi^0$ production \cite{Anisovich_12}. With the available data 
base, they obtain two classes of solutions, called BnGa2011-01 and BnGa2011-02, which differ 
for the resonance contributions in a few partial waves. Predictions for the beam-helicity 
asymmetries from the two solutions are compared to the data in Figs.~\ref{fig:integ_asym} 
(left hand side) and \ref{fig:nucl_par}. The results shown in Fig.~\ref{fig:integ_asym} 
reproduce the measured asymmetries over the whole energy range reasonably well. More details 
emerge from the comparison of the fitted coefficients $A_{i}$ for model results and measured 
asymmetries in Fig.~\ref{fig:nucl_par}. The magnitude and rapid variation of the $A_{1}$ 
coefficient is excellently reproduced by both solutions. The agreement for the $A_{2}$ parameter
is worse. But apart from the second resonance region below masses of 1.6 GeV, solution 01
is clearly preferred. The $A_3$-coefficient is obviously not yet reproduced by any solution.
Results for the neutron target are not yet available from this coupled channel analysis. 

As discussed in \cite{Zehr_12} the reaction model from Fix and Arenh\"ovel (`Two-Pion MAID') 
\cite{Fix_05} reproduced reasonably well the total cross section, the shapes of the invariant 
mass distributions, and the $I^{\odot}$ asymmetries for incident photon energies between 
700 MeV and 800 MeV (agreement was not so good at lower photon energies, where already
the total cross section was underestimated). 
Predictions from it, which are available for the reaction off the proton and off the neutron, 
are confronted with the measured asymmetries in Figs.~\ref{fig:integ_asym} and \ref{fig:nucl_par}.
The experimental results are dominated in the second resonance region ($W<$1.6 GeV) by
a large negative $A_1$ coefficient and this is nicely reflected by the model predictions, which
reproduce quite well the data; for the neutron even better than for the proton.
This seems to indicate that in this range the main reaction mechanisms are understood
and are similar for protons and neutrons, supporting the dominant contribution
from the $D_{13}$(1525) sequential decay. For this energy range we can also compare the results 
to the low-energy partial-wave decomposition from \cite{Kashevarov_12} with the large 
$J=3/2$ component. This is shown in Figs.~\ref{fig:nucl_par} and \ref{fig:integ_asym}
for the $W$ range up to 1575 MeV ((blue) diamonds and dotted curves, respectively).
The agreement with the measured asymmetries is worse than for the Two-Pion MAID model
\cite{Fix_05} and the BnGa analysis.

At higher incident photon energies, between the two nucleon resonance peaks, agreement
between measurements and model predictions for the proton target is still reasonable.
In the third resonance peak above 1.7 GeV it becomes poorer, but Fig.~\ref{fig:nucl_par} 
shows that even here most coefficients follow the tendency of the measured asymmetries.
For the neutron target, measured and predicted asymmetries disagree strongly above $W>$1.65 GeV.

\section{Summary and conclusions}
Precise results have been measured for the beam-helicity asymmetries in the production of
$\pi^0$-pairs off free protons from a hydrogen target and off quasi-free protons and neutrons from
a deuterium target with a circularly polarized photon beam. The measured asymmetries are sizable.
The following are the most important experimental findings. 

The results for the measurements off free and quasi-free protons are identical within statistical 
uncertainties when the kinematics of the quasi-free reaction is reconstructed.
This demonstrates that in the investigated energy range $I^{\odot}$ can be measured 
for $\pi^0$ pairs in quasi-free kinematics off a deuterium target, which is the basis for the measurement 
off quasi-free neutrons. This is far from trivial since the observable $I^{\odot}$ depends in a
subtle way on the correlation of the relative angles of all final state particles, which could be
easily disturbed by Fermi motion or final state interactions. This result is also encouraging
in view of the current experimental program for the measurement of double polarization observables off 
bound neutrons \cite{Krusche_11}, although for each observable and each reaction channel it will be 
necessary to test first the assumption of unmodified quasi-free production for the proton case.  

The comparison of the measured asymmetries and model predictions for photoproduction off the proton 
demonstrates that advanced reaction models like the BnGa-analysis \cite{Anisovich_12} or the
Two-Pion MAID model \cite{Fix_05} can already now make reasonable predictions for this observable.
The outstanding discrepancies will contribute to further refinement of the models.  

The asymmetries measured for the $\vec{\gamma} n\rightarrow n\pi^0\pi^0$ reaction are almost the
same as for the proton target. This could be expected for the second resonance region, where nucleon 
resonance contributions to $\pi^0$ pairs are similar for protons and neutrons. The asymmetries agree 
nicely in this region with the Two-Pion MAID prediction \cite{Fix_05}. The similarities are surprising
at higher incident photon energies, where different nucleon resonances are believed to dominate
the reactions for the proton and the neutron. Model predictions for this energy range do not yet
reproduce the measurements, but will certainly profit when they are included into the fits. 

\newpage
{\bf Acknowledgments}

We wish to acknowledge the outstanding support of the accelerator group 
and operators of MAMI. We thank the members of the Bonn-Gatchina analysis
group (V. Nikonov and U. Thoma) for the provision of the unpublished asymmetries 
from the coupled channel analysis.
This work was supported by Schweizerischer Nationalfonds, Deutsche
Forschungsgemeinschaft (SFB 443, SFB/TR 16), DFG-RFBR (Grant No. 05-02-04014),
UK Science and Technology Facilities Council, STFC, European Community-Research 
Infrastructure Activity (FP6), the US DOE, US NSF and NSERC (Canada).

\end{document}